\documentclass[twocolumn,aps,pra,superscriptaddress,a4paper,sort&compress,balancelastpage]{revtex4-2}
\usepackage[normalem]{ulem}
\usepackage{amsmath}
\usepackage{amssymb}
\usepackage{gensymb}
\usepackage{bm}
\usepackage{graphicx}
\usepackage{float}
\usepackage{subfigure}
\usepackage[dvipsnames]{xcolor}
\usepackage{placeins}
\usepackage{braket}
\usepackage{bbold}
\usepackage{ulem}
\usepackage[colorlinks, linkcolor=blue, citecolor=blue, urlcolor=blue, breaklinks=red]{hyperref}

\usepackage{bbm}
\newcommand{\1}{\mathbbm{1}}
\usepackage{lipsum}
\usepackage{physics}
\usepackage{dsfont}

\definecolor{Mycolor1}{HTML}{44aa99}
\definecolor{Mycolor2}{HTML}{cc6677}
\usepackage{orcidlink}

\begin{document}

\title{Quantum obesity and steering ellipsoids for fermionic fields in dilaton black hole}

\author{Samira Elghaayda~\!\!\orcidlink{0000-0002-6655-0465}}
\email{samira.elghaayda-etu@etu.univh2c.ma}
\affiliation{Laboratory of Mechanics and High Energy Physics, Department of Physics,\\ Faculty of Sciences of Aïn Chock, Hassan II University,\\ P.O. Box 5366 Maarif, Casablanca 20100, Morocco}

\author{M. Y. Abd-Rabbou~\!\!\orcidlink{0000-0003-3197-4724}}
\email{m.elmalky@azhar.edu.eg}
\affiliation{School of Physics, University of Chinese Academy of Science,\\ Yuquan Road 19A, Beijing, 100049, China}
\affiliation{Mathematics Department, Faculty of Science, Al-Azhar University, \\ Nassr City 11884, Cairo, Egypt}

\author{Mostafa Mansour~\!\!\orcidlink{0000-0003-0821-0582}}
\email{mostafa.mansour.fsac@gmail.com}
\affiliation{Laboratory of Mechanics and High Energy Physics, Department of Physics,\\ Faculty of Sciences of Aïn Chock, Hassan II University,\\ P.O. Box 5366 Maarif, Casablanca 20100, Morocco}

\begin{abstract}
This paper investigates quantum obesity (QO), quantum discord (QD), and the quantum steering ellipsoid (QSE) for bipartite Gisin states subjected to Garfinkle-Horowitz-Strominger (GHS) dilation of spacetime on the second qubit. These three quantifiers are introduced to characterize quantum correlations beyond entanglement and can also function as entanglement witnesses. Our results demonstrate a monotonic decrease in the physical accessibility of both QD and QO as the dilation parameter increases within the region-I of the second qubit. Conversely, in the anti-particle region, the accessibility of QD and QO stabilizes at finite values of the dilation parameter owing to the influence of the Pauli exclusion principle and Fermi-Dirac statistics, subsequently increasing gradually. Notably, the QSE in the region-I expand as the Dirac field frequency rises and the dilation parameter diminishes, while the opposite trend is observed in the anti-particle region.
\end{abstract}

\date{\today}

\maketitle


\section{Introduction}
Relativistic quantum field theory (QFT) offers a comprehensive framework for comprehending our Universe, bridging the principles of quantum mechanics and relativistic causality \cite{caribe2023lensing}. It introduces the concept of the quantum vacuum and its inherent fluctuations. Even in the vacuum state of free fields on a flat spacetime, there exists a fundamental entanglement between spatially separated degrees of freedom of a quantum field \cite{summers1985vacuum,summers1987bell}. This vacuum entanglement stems from the interaction between neighboring field oscillators within wave equations, necessary for wave propagation. Such interaction is facilitated through spatial derivatives in operators like the d'Alembertian and Dirac operators. As a result, the ground state of local field oscillators becomes entangled. 

The intersection of relativistic QFT and quantum information has gained considerable interest \cite{fuentes2005alice, adesso2007continuous, shi2018quantum, viermann2022quantum, abd2023detraction, elghaayda2023entropy, wu2024genuinely,2025quantum, elghaayda2024distribution, elghaayda2024physically}, as it not only enhances our understanding of quantum mechanics but also offers new insights into the information paradox associated with black holes (BHs) \cite{hawking1976breakdown}. Typically, BHs are compact objects that contain a massive amount of mass in a very small region of spacetime \cite{carroll2019spacetime,wald2010general}. They have a surface known as the event horizon. Once a particle crosses the event horizon, it is believed that it can never escape, regardless of the energy invested in the process. Initially, it was believed that BHs were completely black and unable to emit particles. However, Hawking subsequently demonstrated that BHs also emit radiation \cite{hawking1975particle, hartle1976path}. This discovery was groundbreaking because, without this radiation emission, certain fundamental laws of thermodynamics would be violated \cite{bekenstein2020black,bekenstein1973black,hawking1972black}.

On the other hand, dilatonic BHs have emerged as an intriguing subject in the study of astrophysical BHs \cite{anabalon2013exact,badia2020influence}. In the context of string theory with finite energy, the solution for a static, spherically symmetric, charged dilatonic BH was derived in \cite{gibbons1988black,garfinkle1991charged}. The dilation parameter $D$ plays a crucial role in determining the intensity of the additional attractive force. When $D = 0$, the BH possesses an inner horizon. The quantum properties of dilatonic BHs, particularly in relation to quantum information and general relativity, are of considerable interest. For instance, He et al. investigated how the dilation parameter $D$ influences measurement-induced non-locality for Dirac particles in GHS dilation spacetime \cite{he2016measurement}. Lian et al. examined the quantum Fisher information of hybrid systems in the context of GHS dilation spacetime \cite{lian}. More recently, Ming et al. explored the non-local advantage of quantum coherence within a GHS dilation spacetime background \cite{ming2024nonlocal}. Notably, studying quantum correlations in GHS dilation spacetime has significant implications for quantum information science \cite{wang2009entanglement}, string theory \cite{garfinkle1991charged, porfyriadis2023charged}, and loop quantum gravity \cite{perez2017black}.

The Bloch sphere provides a straightforward geometric representation for any qubit state \cite{nielsen2010quantum}. However, visualizing a bipartite state geometrically is more complex. Jevtic et al. addressed this challenge by introducing the QSE \cite{jevtic2014quantum}. This tool leverages the concept of EPR steering, where measuring one part of an entangled state affects the other part from a distance. By applying local operations to one qubit, the Bloch vectors of the resulting steered states on the Bloch sphere form a QSE. Milne et al. established the conditions under which a QSE can represent a bipartite state \cite{milne2014quantum}. The QSE not only provides a visual representation but also indicates quantum correlations. Experimental observations of QSE have been made using all-optical photonic qubits \cite{xu2024experimental}. QO, which extends beyond entanglement, is defined through steering processes and QSE \cite{jevtic2014quantum}. It measures quantum correlations in a less restrictive manner than entanglement but more so than QD \cite{milne2014quantum,rosario2024quantum}. QO has been proposed as a metric for analyzing entanglement swapping protocols \cite{rosario2023swapping}. QD, which quantifies the quantum nature of correlations, has led to insights into the effects of decoherence on quantum systems \cite{daoud2012quantum, girolami2011quantum}. Despite its clear conceptual definition, computing QD is challenging due to the need to optimize measurements on a single qubit, though it can be analytically studied for $X$ states \cite{huang2014computing, maldonado2015analytical}.

Building upon the preceding studies, this study delves into the exploration of the QO, QD, and QSE for Dirac particles within the context of the GHS dilation of spacetime. The primary objective is to identify patterns in the distribution of QO and QD as a method of assessing quantum correlation. Additionally, we will explore the QSE. To illustrate this, we consider a scenario where two observers, Alice and Bob, initially share a Gisin state in flat Minkowski spacetime. Alice remains stationary in the asymptotically flat region, while Bob free-falls towards a dilating black hole, eventually approaching the event horizon. Within this setup, we conduct an in-depth analysis of the three quantum quantifiers to uncover key properties under different system parameters.

The remainder of this paper is structured as follows: Sec. \ref{sec2} introduces the fundamental concepts utilized in our study to quantify quantum correlations, including QO, QSE, and QD. We carry out these quantifications within the GHS dilation spacetime framework, where Alice and Bob share a Gisin state. In Sec. \ref{model}, we provide an in-depth review of the vacuum structure and the Dirac field quantization within this GHS dilation spacetime. Sec. \ref{sec4} details the key results derived from our investigation. Finally, Sec. \ref{sec4} concludes with a brief summary. For simplicity, \(G\), \(\hbar\), \(c\), and \(k_B\) are set to unity throughout the paper.
\section{Quantum metrics} \label{sec2}
In this section, we will provide a concise overview of QD and QO, which are  employed to assess quantum correlations. Furthermore, we will explore QSE, a natural extension of the Bloch ball model that offers fresh perspectives on quantum entanglement and steering.
\subsection{QO for bipartite states}
We present the general formulation of QO and offer several useful analytical solutions for various families of quantum states. A quantum state can be described using the Bloch representation \cite{nielsen2010quantum}
\begin{align}
\hat{\eta}^{\text{AB}}=\frac{1}{4}\sum_{i=0}^{3}\sum_{j=0}^{3}\mathcal{R}_{ij}\hat{s}_{i}^{\text{A}}\otimes \hat{s}^{\text{B}}_{j} , \label{eq:General_state}
\end{align}
with $\hat{s}_{0}=1$ and $\{\hat{s}_{i}\}_{i=1}^{3}$ representing the Pauli operators, we can uncover the correlations both quantum and classical of a bipartite state by analyzing the matrix $\mathcal{R}$. This matrix comprises the elements $\mathcal{R}_{ij} = \langle \hat{s}_{i} \otimes \hat{s}_{j} \rangle$ \cite{gamel2016entangled}, which are given by
\begin{align}
\mathcal{R}=\begin{pmatrix}
1 & v_{1} & v_{2}& v_{3}\\
w_{1} &\Theta_{11}&\Theta_{12}&\Theta_{13} \\
w_{2}&\Theta_{21}  & \Theta_{22}& \Theta_{23}\\
w_{3}& \Theta_{31}  & \Theta_{32} & \Theta_{33}
\end{pmatrix}=\begin{pmatrix}
1 & \vec{v}\\
\vec{w}^{T} &\Theta
\end{pmatrix} .
\end{align}
with $\vec{v} = [v_{1}, v_{2}, v_{3}]$ and $\vec{w} = [w_{1}, w_{2}, w_{3}]$ denote the local Bloch vectors for part A and part B, respectively. Then we have 
\begin{subequations}\label{eq:BlochVectors}
\begin{align}
\vec{v} &= [ \mathrm{tr}(\hat{\eta} \hat{s}_{1}^{A}\otimes\hat{\1}^{B}), \mathrm{tr}(\hat{\eta} \hat{s}_{2}^{A}\otimes\hat{\1}^{B}), \mathrm{tr}(\hat{\eta} \hat{s}_{3}^{A}\otimes\hat{\1}^{B})] ,  \\
\vec{w} &= [ \mathrm{tr}(\hat{\eta} \hat{\1}^{A}\otimes\hat{s}_{1}^{B}), \mathrm{tr}(\hat{\eta} \hat{\1}^{A}\otimes\hat{s}_{2}^{B}), \mathrm{tr}(\hat{\eta} \hat{\1}^{A}\otimes\hat{s}_{3}^{B})] .
\end{align} 
\end{subequations}
The block matrix $\Theta$ is significant because its singular values are essential for computing quantum correlations \cite{horodecki1995violating,costa2016quantification,abd2022improving}. Additionally, the correlation matrix $\mathcal{R}$ is used to define the QO for bipartite systems within the Hilbert space $\mathcal{H}_{2} = \mathbf{C}^{2}\otimes \mathbf{C}^{2}$. The QO is formally characterized by the determinant of the matrix $\mathcal{R}$ \cite{milne2014quantum,rosario2024quantum}.
\begin{align}
\Omega=\vert\mathrm{det}(\mathcal{R})\vert^{1/4} .
\label{eq:Obesity}
\end{align}
while the QO can be computed analytically, we will concentrate on three families of states that are particularly relevant to the phenomenon under investigation in this work for practical purposes. The states of interest can be uniformly expressed as
\begin{align}
\hat{\eta}_{(k)} =\begin{pmatrix}
\eta_{11} & \delta_{k3}\eta_{12} & \delta_{k2}\eta_{13} & \eta_{14}\\
\delta_{k3}\eta_{21} &\eta_{22}&\eta_{23}& \delta_{k2}\eta_{24} \\
\delta_{k2}\eta_{31}&\eta_{32}  & \eta_{33} & \delta_{k3}\eta_{34}\\
\eta_{41}& \delta_{k2}\eta_{42}  & \delta_{k3}\eta_{43} & \eta_{44}
\end{pmatrix} , \label{eq:DensityMatrixGeneric}
\end{align}
where $\delta_{kn}$ denotes the Kronecker delta and $k = \{1, 2, 3\}$. It is important to note that, apart from the state $\hat{\eta}_{(1)}$, the states $\hat{\eta}_{(k)}$ cannot be represented as $X$-states. For this specific set of states, the QO simplifies
\begin{align}
\Omega=2\abs{ \big(|\eta_{23}|^{2}-|\eta_{14}|^{2}\big)\big(\eta_{22}\eta_{33}-\eta_{11}\eta_{44}\big)}^{1/4} .
\label{eq:Obesity_X}
\end{align}
This result will be pivotal throughout our manuscript for analyzing quantum correlations in Dirac particles within a GHS dilation spacetime background. The geometric interpretation of quantum observables can be understood by examining the steering ellipsoid associated with the quantum state under consideration. Let $\hat{\eta}^{\text{AB}}$ denote a general bipartite state as specified in Eq.~\eqref{eq:General_state}. In this context, local measurements and classical communication (LOCC) in subsystem B will "steer" the Bloch sphere of subsystem A to a new geometry represented by an ellipsoid \cite{jevtic2014quantum}. Specifically, $\hat{\eta}^{\text{A}}_{\text{steer}} = \text{Tr}_{B}(\hat{\eta}^{\text{AB}}_{\text{LOCC}})$ denotes the reduced density matrix of subsystem A after local measurements in B. Here, $\hat{\eta}^{\text{AB}}_{\text{LOCC}} = \text{LOCC}[\hat{\eta}^{\text{AB}}]$ represents the density matrix post-LOCC in subsystem B. Consequently, any state $\hat{\eta}^{\text{A}}_{\text{steer}}$ can be visualized as residing within an ellipsoidal region in the Bloch sphere of A. This ellipsoidal region is characterized by a vector $\vec{c}_{A}$ and a matrix $\mathcal{Q}_{A}$, defined respectively by
\begin{align}
\vec{c}_{A}&= \gamma_{b} \left(\vec{v}-\Theta\cdot\vec{w}\right) ,\\
\mathcal{Q}_{A}&=\gamma_{b}\left(\Theta-\vec{v}\cdot\vec{w}^{T}\right)\left(\1 +\gamma_{b}\vec{w}\cdot\vec{w}^{T}\right)\left(\Theta^{T}-\vec{w}\cdot\vec{v}^{T}\right),
\end{align}
where $\vec{c}_{A}$ denotes the center of an ellipsoid within the local Bloch sphere of subspace $\text{A}$. The semi-axis lengths $s_{i}=\sqrt{q_{i}}$ and the orientation of the ellipsoid are defined by the eigenvalues $\{q_{i}\}_{i=1}^{3}$ and eigenvectors of $\mathcal{Q}_{A}$, respectively. The parameter $\gamma_{b}$ is given by $1/(1-|\vec{w}|^{2})$, with the assumption that $|\vec{w}|<1$. By incorporating both the vector component $\vec{c}_{A}$ and the semi-axis lengths $s_{i}$, Alice's QSE, as influenced by Bob, can be determined by
\begin{eqnarray}
	\varepsilon_{A}=\left\{|\left(
	\begin{array}{ccc}
	c_{A}(1)\\
	c_{A}(2)\\
	c_{A}(3)\\
	\end{array}
	\right)+\left(
	\begin{array}{ccc}
	s_{1} x\\
	s_{2} y\\
	s_{3} z\\
	\end{array}
	\right) | x\leq1 \right\},
	\end{eqnarray} 

\subsection{QD for bipartite states}
We define discord as a metric for the quantumness of correlations \cite{daoud2012quantum,girolami2011quantum}. It is calculated as the difference between the total and classical correlations in the system. $X$ states represent the most comprehensive class of quantum states that can be analytically examined with respect to QD. Specifically, for a bipartite $X$ state $\hat{\eta}^{\text{AB}}$, the QD is mathematically defined as \cite{fanchini2010non,ding2011quantum}.
\begin{equation}
\mathcal{D}(\hat{\eta}^{\text{AB}})=\min(\mathcal{O}_1,\mathcal{O}_2),
\end{equation}
where
\begin{equation}
\mathcal{O}_k=S(\hat{\eta}^{\text{B}})-S(\hat{\eta}^{\text{AB}})-H_k,
\end{equation}
given the von Neumann entropy of the general density matrix $\hat{\eta}^{\text{AB}}$ is expressed as $S(\hat{\eta}^{\text{AB}}) =- \mathrm{tr}\left[ \hat{\eta}^{\text{AB}} \log_2 \hat{\eta}^{\text{AB}} \right]$, we can similarly define $S(\hat{\eta}^{\text{B}}) = -\mathrm{tr}\left[ \hat{\eta}^{\text{B}} \log_2 \hat{\eta}^{\text{B}} \right]$ as the von Neumann entropy of the reduced density matrix for the second qubit (labeled $B$), upon which the measurement is performed. The quantities $H_{k=1,2}$ are defined by
\begin{equation}
\begin{split}
&H_1=\sum_{i=1}^{2}\left\lbrace (\iota \frac{1+(-1)^i\beta}{2} \log_2 \frac{1+(-1)^i\beta}{2}\right.  \\&\left.+\epsilon\frac{1+(-1)^i\tau}{2} \log_2 \frac{1+(-1)^i\tau}{2}\right\rbrace, \\&
H_2=\sum_{j=1}^{2}\left\lbrace (\frac{1+(-1)^j\varsigma}{2} \log_2 \frac{1+(-1)^j\varsigma}{2}\right\rbrace,
\end{split}
\end{equation}
where $\iota= \eta_{11}+\eta_{33}$,  $\epsilon=\eta_{22}+\eta_{44}$, $\beta=\frac{|\eta_{11}-\eta_{33}|}{\eta_{11}+\eta_{33}}$, $\tau=\frac{|\eta_{22}-\eta_{44}|}{\eta_{22}+\eta_{44}}$, and $\varsigma= \sqrt{(\eta_{11}+\eta_{22}-\eta_{33}-\eta_{44})^2+4(|\eta_{14}|+|\eta_{23}|)^2}$, while $\eta_{ij}$ represent the elements of the density operator $\hat{\eta}^{\text{AB}}$. 

\section{Gisin state in the GHS dilation spacetime \label{model}}
In this section, our objective is to explore the concepts of fermionic fields and the vacuum structure in GHS dilation spacetime. We will focus on the viewpoint of observers who are initially in a Gisin state. It is worth mentioning that this state exhibits full entanglement for an observer who is in free fall towards a GHS dilation spacetime.

\subsection{Vacuum Structure of fermionic fields in GHS dilation spacetime}
\label{sec3}
In this part, we will provide a brief overview of the treatment of a quantum fermionic field within GHS dilation spacetime. Additionally, we will examine the vacuum structure of a Dirac particle, which is essential for comprehending the interaction between the field and the observer. Our focus will specifically be on the spherically symmetric line element of the GHS dilation spacetime, expressed as follows \cite{ garcia1995class,ming2019dynamical,he2016measurement}
\begin{equation}
ds^2 = r\mathfrak{D} \left(d\theta^2 + \sin^2\theta \, d\Phi^2\right)-\frac{\mathfrak{D}}{\mathfrak{m}} dt^2 +(\frac{\mathfrak{m}}{\mathfrak{D}})^{-1} dr^2 
\end{equation}
where $\mathfrak{D}=r - 2D$, and $\mathfrak{m}=r - 2M$. We denote \( M \) as the BH mass and \( D \) as the field's dilation parameter. The relationship between \( M \), the charge \( Q \), and \( D \) is given by \( D = \frac{Q^2}{2M} \). In this spacetime, the Dirac equation can be derived as follows
\begin{equation}
e^\mu_a \gamma^a (\partial_\mu + \Gamma_\mu) \Upsilon = 0
\end{equation}
where \(\Gamma_\mu\) denotes the spin connection coefficient, \(e^\mu_a\) represents the inverse of the tetrad \(e^a_\mu\), and \(\gamma^a\) signifies the Dirac matrices. We introduce a tortoise coordinate
\begin{equation}
r^* = 2\Lambda \ln \left[ \frac{\mathfrak{m}}{2\Lambda} \right] + r 
\end{equation}
where $\Lambda = M - D$, we define the advanced time by $\upsilon = t + r^*$ and the retarded time by $u = t - r^*$. By solving the Dirac equation in the GHS dilation spacetime, we can easily obtain the positive frequency (fermion) outgoing solutions both inside and outside the event horizon. These solutions are given by $\Upsilon^{I+}_{\vec{q}} = \mathcal{P} e^{-i\omega_i u}$ and $\Upsilon^{II+}_{\vec{q}} = \mathcal{P} e^{i\omega_i u}$, where $\mathcal{P}$ represents a 4-component Dirac field and $\omega_i$ is the monochromatic frequency of the Dirac field. It is clear that $\Upsilon^{I+}_{\vec{q}}$ and $\Upsilon^{II+}_{\vec{q}}$ form a set of orthogonal basis functions. Consequently, we can expand the Dirac field $\Upsilon_{\text{out}}$ as
\begin{equation}\label{eq9}
\Upsilon_{\text{out}} = \sum_{i, \kappa} \int dq \left( a_{\vec{q}}^{\kappa} \Upsilon_{\vec{q}}^{\kappa^+} + b_{\vec{q}}^{\kappa^*} \Upsilon_{\vec{q}}^{\kappa^-}  \right),
\end{equation}
where $\kappa = (I, II)$ denotes the indices where $a_{\vec{q}}^I$ and $b_{\vec{q}}^{I*}$ correspond to the annihilation operator for fermions and the creation operator for anti-fermions, respectively, applied to the states in the exterior region. Additionally, the generalized light-like Kruskal coordinates $U$ and $V$ for the GHS dilaton spacetime can be represented as follows
\begin{equation}
\nu= 4\Lambda \ln \left( \frac{V}{4\Lambda} \right), \quad u=-4\Lambda \ln \left( \frac{-U}{4\Lambda} \right), \quad \text{if}\quad r>r_{+}, 
	\label{eq11}
\end{equation}
and
\begin{equation}
\nu=4\Lambda \ln \left( \frac{V}{4\Lambda} \right), \quad u= -4\Lambda \ln \left( \frac{U}{4\Lambda} \right), \quad \text{if}\quad r<r_{+}, 
	\label{eq12}
\end{equation}
By extending $\Upsilon^{I+}_{\vec{q}}$ and $\Upsilon^{II+}_{\vec{q}}$ analytically, we can derive complete bases for the positive-energy modes for all real values of $U$ and $V$. This process utilizes the connection between Kruskal coordinates and BH coordinates, as proposed by Damour and Ruffini.
\begin{equation}
\begin{aligned}
    \delta^{I+}_{\vec{q}} &=\Upsilon^{I+}_{\vec{q}} e^{2\Lambda \pi \omega_{i}}  +\Upsilon^{II+}_{\vec{-k}} e^{-2\Lambda \pi \omega_{i}},  \\
    \delta^{II+}_{\vec{q}} &=\Upsilon^{I-}_{-\vec{q}} e^{-2\Lambda \pi \omega_{i}}  + \Upsilon^{II+}_{\vec{q}}e^{2\Lambda \pi \omega_{i}}.
\end{aligned}
\end{equation}
Consequently, in the framework of Kruskal spacetime, the Dirac fields are expressed as follows
\begin{equation}\label{eq12}
    \Upsilon_{\text{out}} = \sum_{i,\kappa} \int d\vec{q} \left[2 \cosh (4\Lambda \pi \omega_{i}) \right]^{-1/2} \left( l^{\kappa}_{\vec{q}} \delta^{\kappa^+}_{\vec{q}} + \vartheta^{\kappa^*}_{\vec{q}} \delta^{\kappa^-}_{\vec{q}}  \right).
\end{equation}
where $\vartheta^{\kappa^*}_{\vec{q}}$ and $l^{\kappa}_{\vec{q}}$ denote the creation and annihilation operators corresponding to the Kruskal vacuum. Eqs \eqref{eq9} and \eqref{eq12} represent the decompositions of the Dirac fields into GHS dilation and Kruskal modes, respectively. Consequently, suitable Bogoliubov transformations can be formulated between the creation and annihilation operators in the GHS dilation and Kruskal coordinates. Each annihilation operator $l_{\vec{q}}^I$ can be expressed as a combination of the dilation particle operators with a single frequency $\omega_i$ and the orthonormal modes.
\begin{equation}
l_{\vec{q}}^I=a_{\vec{q}}^I\varepsilon_1 - b_{\vec{q}}^{II^*}\varepsilon_2
\end{equation}
we define \(\varepsilon_1 = \left[1 + e^{-8\Lambda\pi w_i}\right]^{-1/2}\) and \(\varepsilon_2 = \left[1 + e^{8\Lambda\pi w_i}\right]^{-1/2}\). The GHS dilaton spacetime is divided into two regions: the physically accessible region \(I\) and the inaccessible region \(II\). In the GHS dilaton spacetime coordinates, the ground-state mode is mapped into a two-mode squeezed state under Kruskal coordinates. By normalizing the state vectors, the Kruskal-particle vacuum state for the mode \(\vec{q}\) is described as follows
\begin{equation}\label{eqk}
|0_{\vec{q}} \rangle^+_K = \varepsilon_1 |0_{\vec{q}} \rangle^+_I |0_{-\vec{q}}\rangle^-_{II} +\varepsilon_2 |1_{\vec{q}} \rangle^+_I |1_{-\vec{q}}\rangle^-_{II}.
\end{equation}
the states $\{|m_{\pm\vec{q}} \rangle_{I,II}^\pm\}$ form an orthonormal basis corresponding to the regions outside and inside the event horizon, respectively. The superscript $\{\pm\}$ on the kets indicates the particle and antiparticle vacua. In a similar manner, we can derive the one-excitation state.
\begin{equation}\label{eqkk}
|1_{\vec{q}} \rangle_{K}^+ = |1_{\vec{q}} \rangle_{I}^+ |0_{-\vec{q}} \rangle_{II}^- 
\end{equation}
From now on, we assume that \(\{|n_{\vec{q}} \rangle_{I}^+\}\) as \(\{|n \rangle_{I}\}\) and \(\{|n_{-{\vec{q}}} \rangle_{II}^-\}\) as \(\{|n \rangle_{II}\}\) for brevity. Additionally, we take \(\omega_i = \omega = 1\) for simplicity.
\subsection{Derivation of the evolved Gisin state}
Gisin states, introduced by Gisin in \cite{Gisin1996hidden}, represent a class of states notable for their unique properties. These states encompass combinations of pure, entangled, and separable mixed states, and have been instrumental in demonstrating hidden nonlocality. Specifically, certain Gisin states initially exhibit local behavior but can lose this characteristic after undergoing purely local operations \cite{Gisin1996hidden, gs2,elghaayda2023p}. Mathematically, the Gisin states can be expressed as:
\begin{equation}
	\eta_{g,\alpha}= g | \xi_{\alpha}\rangle \langle \xi_{\alpha}| + \frac{1 - g}{2} \left( |00\rangle\langle00| + |11\rangle\langle 11| \right),
	\label{eq19}
\end{equation}
The mixing parameter $g \in [0, 1]$ defines the degree of mixture, while the state $|\xi_{\alpha}\rangle = \sin(\alpha)|01\rangle + \cos(\alpha)|10\rangle$ represents a Bell-like state with $\alpha \in [0, \pi/2]$. At $\alpha = \frac{\pi}{4}$, $|\xi_{\alpha}\rangle$ becomes the familiar Bell state $|\xi^{+}\rangle = \frac{1}{\sqrt{2}}(|01\rangle + |10\rangle)$. Gisin states are distinct from Weyl states but intersect non-trivially when locally maximally mixed at $\alpha = \frac{\pi}{4}$. For $g < 1$, the Gisin state is fully mixed. However, at $g = 1$, it simplifies to the pure entangled state $\left| \xi_{\alpha}\right\rangle \left\langle \xi_{\alpha}\right|$.

\begin{figure}[H]
	\centering
	\includegraphics[width=0.9\linewidth]{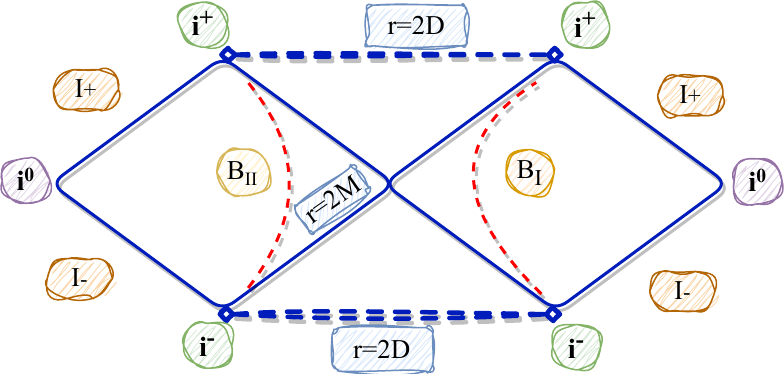}
	\caption{The Penrose diagrams for the GHS dilation BH depict the world-lines of Bob in the particle and anti-particle regions. In this illustration, $i^0$ corresponds to spatial infinities, while $i^+$ and $i^-$ correspond to time-like past and future infinity, respectively. $I_+$ and $I_-$ represent light-like past and future infinity.}
	\label{fig:b}
\end{figure}
Consider that Alice and Bob initially share a Gisin state $\eta_{g,\alpha}$ in flat Minkowski spacetime. Alice's detector measures mode $\left|n\right\rangle_A$, while Bob's detector measures mode $\left|n\right\rangle_B$. Alice remains in the asymptotically flat region, while Bob falls towards a GHS dilation spacetime and hovers near the event horizon, as shown in FIG. (\ref{fig:b}). As Bob traverses the Kruskal vacuum, his detector measures a thermal Fermi–Dirac distribution of particles due to the Hawking effect. To describe what Bob observes in this curved spacetime, the mode $\left|n\right\rangle_B$ must be specified in the coordinates of the GHS dilation spacetime. The Hawking temperature that Bob experiences is given by $T=1/8\pi \Lambda$. To describe Bob's perspective, the initial Gisin state corresponding to mode $q$ must be expanded, considering both the interior and exterior regions of the event horizon. The Kruskal states $\left|0 \right\rangle_{K}$ and $\left|1 \right\rangle_{K}$ correspond to two-mode states within the GHS dilation spacetime, as described by Eqs \eqref{eqk} and \eqref{eqkk}. By rewriting the initial Gisin state, the evolved state can be expressed as
\begin{widetext}
\begin{minipage}{\linewidth}
\begin{eqnarray}
	\eta_{AB_{I}B_{II}} &=& \frac{1-g}{2}\left( \varepsilon_1^{2} |000\rangle\langle000|+\varepsilon_1\varepsilon_2(|000\rangle\langle011|+|011\rangle\langle000|)+\varepsilon_2^{2}|011\rangle\langle011|+|110\rangle\langle110|\right) \nonumber\\&+&g\cos^{2}\alpha\left(\varepsilon_1^{2} |100\rangle\langle100|+\varepsilon_1\varepsilon_2(|100\rangle\langle111|+|111\rangle\langle100|+\varepsilon_2^{2}|111\rangle\langle111|) \right)+ g\sin^{2}\alpha|010\rangle\langle010|\nonumber\\&+&g\sin\alpha\cos\alpha\left(\varepsilon_1(|010\rangle\langle100|+|100\rangle\langle010|)+\varepsilon_2(|010\rangle\langle111|+|111\rangle\langle010|) \right). 
	\label{eq20}
\end{eqnarray}
\end{minipage}
\end{widetext}
By decomposing the density operator $\eta_{AB_{I}B_{II}}$ into one degree of freedom, we can identify three different subsystems: a physically accessible subsystem \(\eta_{AB_{I}}\), a physically inaccessible subsystem \(\eta_{AB_{II}}\), and a spacetime subsystem \(\eta_{B_{I}B_{II}}\).

\section{Results and Discussion \label{sec4}}
Following the above setups, we now study the QD, QO, and QSE of Dirac fields that are present in every subsystem. Although the established global quantum state consists of fermionic modes both inside and outside the event horizon, it is a pure state. However, due to the causal disconnection between the interior and exterior regions, an observer or detector outside the GHS dilation BH cannot access all the information. By using the partial trace operation, one can distinguish between physically accessible and inaccessible scenarios. Despite the inaccessibility of certain scenarios to experimental observation or detection, the global quantum state remains unitary and pure. Now in favour of the conservation of quantum information, we can theoretically investigate both accessible and inaccessible scenarios. The following subsections will explore these scenarios in detail.

\subsection{Bob's qubit in Region-I}
As previously mentioned, the exterior and interior regions of GHS dilation are completely disconnected. Therefore, the information can only be accessed within the subsystem $\eta_{AB_{I}}$. Within this subsystem, the corresponding reduced density operator is as follows:

\begin{eqnarray}
	\eta_{AB_{I}} &=& \frac{1-g}{2}\varepsilon_1^{2} |00\rangle\langle00|+(\frac{1-g}{2}\varepsilon_2^{2}+g\sin^{2}\alpha)|01\rangle\langle01|\nonumber\\&+&g\sin\alpha\cos\alpha\varepsilon_1(|01\rangle\langle10|+|10\rangle\langle01|)\\&+&\nonumber g\cos^{2}\alpha\varepsilon_1^{2}|10\rangle\langle10|+(\frac{1-g}{2}+g\cos^{2}\alpha\varepsilon_2^{2})|11\rangle\langle11|.
	\label{eq21}
\end{eqnarray}

\begin{widetext}
\begin{minipage}{\linewidth}
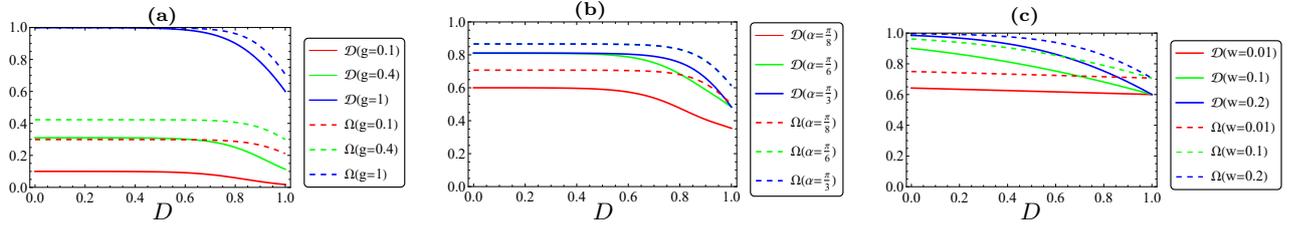
\begin{figure}[H]
	\centering
	\includegraphics[scale=0.42,trim=00 00 00 00, clip]{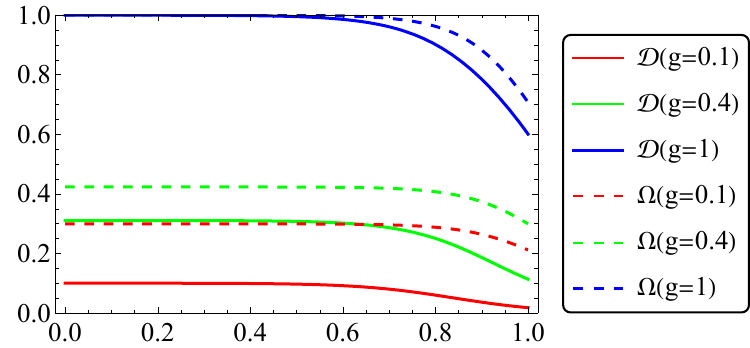} \quad
	\includegraphics[scale=0.42,trim=00 00 00 00, clip]{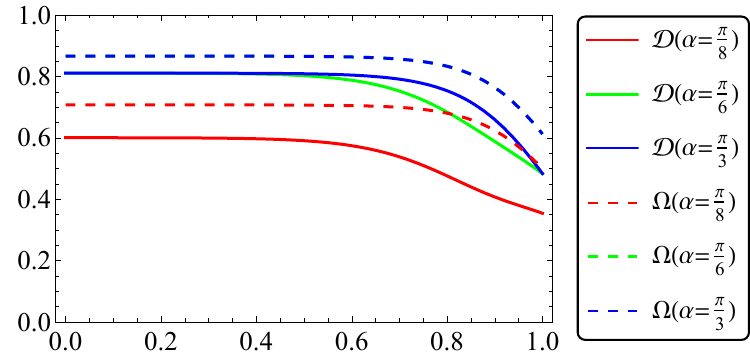} \quad
	\includegraphics[scale=0.42,trim=00 00 00 00, clip]{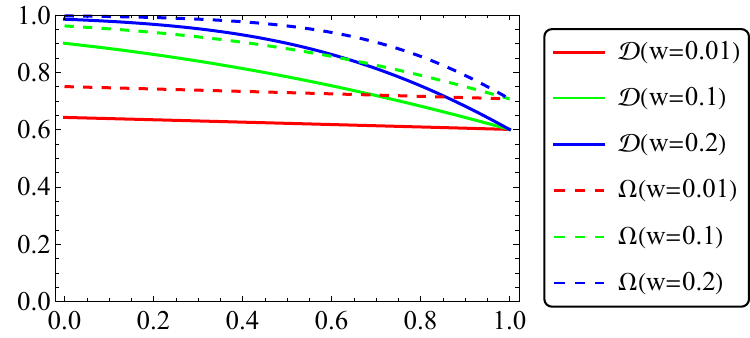} 
	\put(-425,70){$\textbf{{\scriptsize (a)}}$}
 \put(-265,72){$\textbf{{\scriptsize (b)}}$} \put(-105,70){ $\textbf{{\scriptsize (c)}}$} \put(-423,-5){${\scriptsize D}$} \put(-258,-5){${\scriptsize D}$} \put(-98,-5){${\scriptsize D}$} 
	\caption{QD $\mathcal{D}$ (solid-curves), and QO $\Omega$ (dashed-curve) for $\eta_{A B_{I}}$ with (a) $\alpha=\frac{\pi}{4}$, $w=0.5$,  (b) $g=1$, $w=0.5$, (c) $\alpha=\frac{\pi}{4}$, $g=1$.}
	\label{Fig1}
\end{figure}
\end{minipage}
\end{widetext}
In FIG. \ref{Fig1}, we examine the effects of the initial state setting parameters ($g$) and ($\alpha$) as well as the Dirac field frequency on the overall behavior of QO and QD as functions of the GHS dilation parameter ($D$). A general trend observed is that an increase in the dilation parameter correlates with a decrease in the maximum values of both QO and QD. FIG. \ref{Fig1} (a). displays the impact of varying the mixing setting parameter values $g= 0.1,\ 0.4,\ \text{and}\ 1$ with fixed $\alpha=\frac{\pi}{4}$ and $w=0.5$ on QO and QD. It is evident that as $g$ approaches zero, the behavior of both functions converges towards classical states, wherein the maximum bounds of QO and QD diminish to zero. Additionally, at $g=1$, the two functions exhibit identical behavior. However, as $g$ decreases, the disparity between QO and QD functions increases, consistently showing that the maximum bounds of QO are greater than those of QD. FIG. \ref{Fig1}. (b) investigates the influence of the initial state angle ($\alpha = \frac{\pi}{8}, \frac{\pi}{6},$ and $\frac{\pi}{3}$) on QO and QD, while maintaining $g=1$ and $w=0.5$. For $\Omega(\alpha=\frac{\pi}{6})$ and $\Omega(\alpha=\frac{\pi}{3})$, identical patterns are observed, until in the presence of an increase in the dilation parameter $D$. Similarly, $\Omega(\alpha=\frac{\pi}{6})$ and $\Omega(\alpha=\frac{\pi}{3})$ demonstrate analogous behavior, but diverges with increasing $D$. A reduction in the angle to $\alpha = \frac{\pi}{8}$ results in a decrease in the maximum values for both functions. FIG. \ref{Fig1}. (c) explores the impact of varying Dirac field frequency $w= 0.01,\ 0.1, \ 0.2 $ on QO and QD. An increase in frequency is associated with an enhancement in the maximum values of both functions. However, the dilation parameter counteracts this effect, leading to a convergence towards residual values of $0.75 $ for QO and $0.6$ for QD at the maximum dilation.

So, FIG. \ref{Fig1}. shows that an elevated dilation parameter corresponds to a diminished quantum correlation within the region-$ I$. The mixed state component $|00\rangle\langle00| + |11\rangle\langle 11|$ in equation (\ref{eq19}) signifies a decoherence state. A heightened mixing parameter inversely correlates with quantum correlation. The interplay between the initial state angle and the mixing parameter collectively influences the augmentation or reduction of quantum correlation. An increased Dirac frequency enhances quantum correlation, while a constraint emerges for QD and QO at distinct frequencies and elevated dilation values. 

\begin{widetext}
\begin{minipage}{\linewidth}
\begin{figure}[H]
	\centering
	\includegraphics[scale=0.42,trim=00 00 00 00, clip]{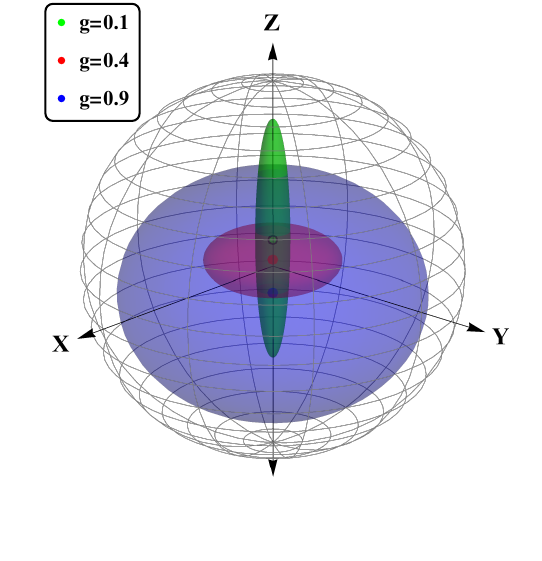} \quad
	\includegraphics[scale=0.42,trim=00 00 00 00, clip]{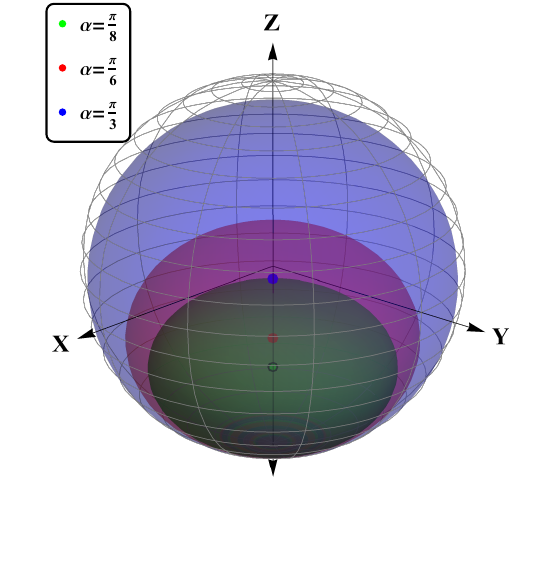} \quad
	\includegraphics[scale=0.42,trim=00 00 00 00, clip]{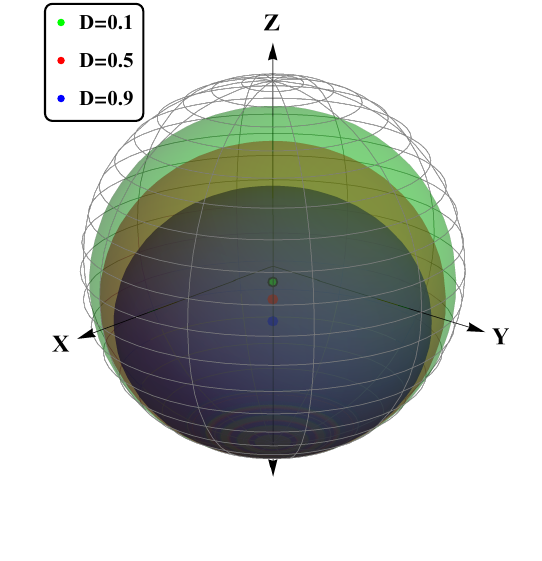} \quad
	\includegraphics[scale=0.42,trim=00 00 00 00, clip]{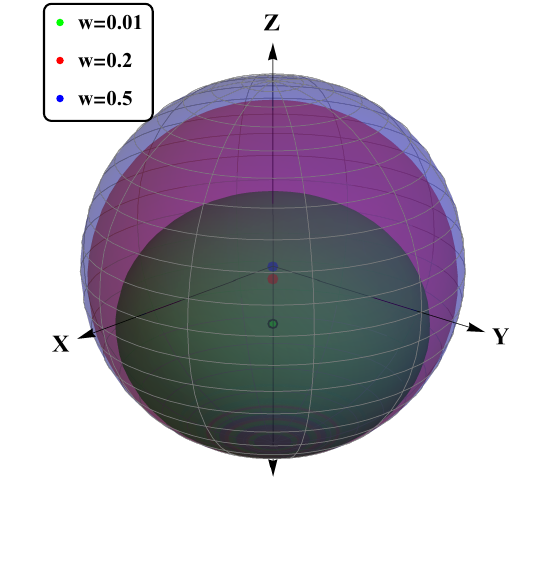} 
	\put(-445,105){$\textbf{{\scriptsize (a)}}$} \put(-325,105){$\textbf{{\scriptsize (b)}}$} \put(-200,105){ $\textbf{{\scriptsize (c)}}$
	}\put(-80,105){ $\textbf{{\scriptsize (d)}}$}
	\caption{Alice’s QSE $\varepsilon_{A}$  corresponding to $\eta_{AB_{I}}$ with fixed $M=1$. (a) $D=0.5$, $w=0.1$, $\alpha=\frac{\pi}{4}$, (b) $g=1$, $w=0.1$, $D=0.5$, (c) $g=1$, $w=0.1$, $\alpha=\frac{\pi}{4}$, (d) $g=1$, $\alpha=\frac{\pi}{4}$, $D=0.5$.}
	\label{Fig2}
\end{figure}
\end{minipage}
\end{widetext}
FIG. \ref{Fig2} proceeds to examine the potential for Alice to steer the quantum state of Bob in the context of QSE for $\eta_{AB_{I}}$ and fixed $M=1$. FIG. \ref{Fig2}. (a) illustrates the impact of varying g values on the QSE. A constraint is observed along the x, y, and z axes for $g = 0.9$, with the ellipsoid's center displaced towards the negative z-axis. For $g = 0.5$, a horizontally oriented ellipsoid emerges, exhibiting greater compression along the z-axis. Conversely, for $g = 0.1$, the ellipsoid is compressed within the x-y plane and its center shifts towards the positive z-axis. FIG. \ref{Fig2} (b) presents that increasing $\alpha$ from $\frac{\pi}{8}$ to $\frac{\pi}{6}$ and subsequently to $\frac{\pi}{3}$ results in an expansion of the ellipsoid volume. Concurrently, the ellipsoid centers converge towards the origin with increasing $\alpha$. In contrast, while an elevated dilation parameter diminishes the ellipsoid volume, it does not lead to complete collapse, as evidenced in FIG. \ref{Fig2}. (c). This implies the feasibility of steering for Alice. FIG. \ref{Fig2}. (d)  reveals a correspondence between the Bloch sphere and the QS ellipsoid at w = 0.5, indicative of substantial quantum correlations and the likelihood of steering.

\subsection{Bob's qubit in Region-II}
We will now investigate how Hawking decoherence affects the properties of the shared Gisin state in the physically inaccessible subsystem $\eta_{AB_{II}}$. To comprehend the whereabouts of the lost quantum correlations, we assume that these vanished quantum correlations are redistributed to Region-II. This assumption is based on the fact that Hawking decoherence splits Bob's mode into mode $\left| n\right\rangle_{B_{I}}$ and mode $\left| n\right\rangle_{B_{II}}$. Our focus will be on the remaining two subsystems, where we will explore the quantum correlations in Region-II. The subsystem $\eta_{AB_{II}}$ is given by

\begin{eqnarray}
	\eta_{AB_{II}} &=&(\frac{1-g}{2}\varepsilon_1^{2}+g\sin^{2}\alpha)|00\rangle\langle00|+ \frac{1-g}{2}\varepsilon_2^{2} |01\rangle\langle01|\nonumber\\&+&g(\frac{1-g}{2}+g\cos^{2}\alpha\varepsilon_1^{2})|10\rangle\langle10|+ g\cos^{2}\alpha\varepsilon_2^{2}|11\rangle\langle11|\nonumber\\&+&g\sin\alpha\cos\alpha\varepsilon_2(|00\rangle\langle11|+|11\rangle\langle00|).
	\label{eq22}
\end{eqnarray}

\begin{widetext}
\begin{minipage}{\linewidth}
\begin{figure}[H]
	\centering
	\includegraphics[scale=0.42,trim=00 00 00 00, clip]{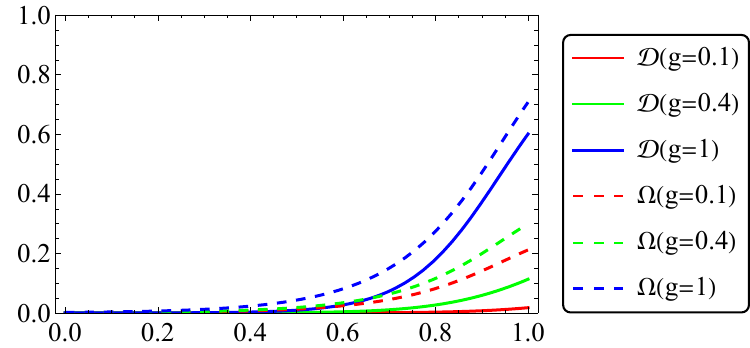} \quad
	\includegraphics[scale=0.42,trim=00 00 00 00, clip]{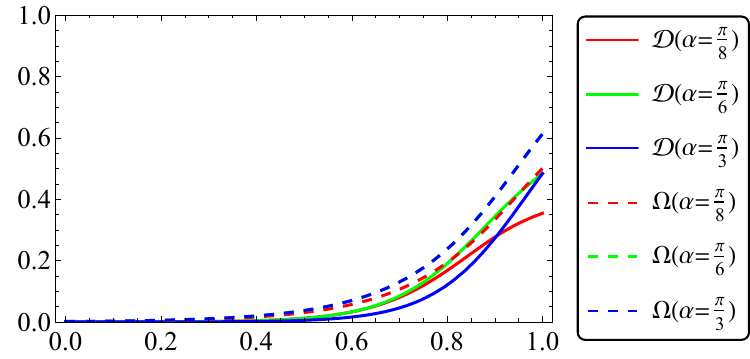} \quad
	\includegraphics[scale=0.42,trim=00 00 00 00, clip]{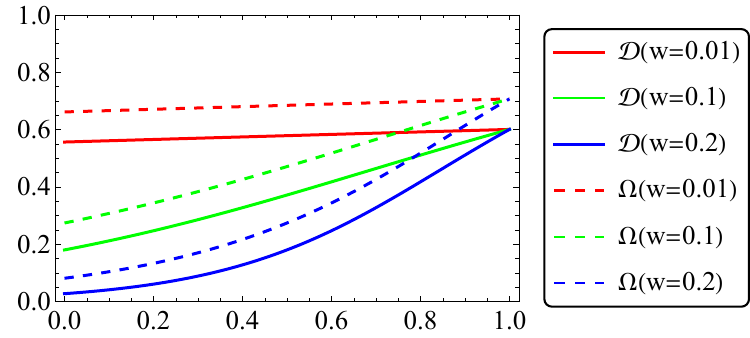} 
	\put(-430,70){$\textbf{{\scriptsize (a)}}$} \put(-265,72){$\textbf{{\scriptsize (b)}}$} \put(-105,70){ $\textbf{{\scriptsize (c)}}$}  \put(-423,-5){${\scriptsize D}$} \put(-258,-5){${\scriptsize D}$} \put(-98,-5){${\scriptsize D}$} 
	\caption{ QD $\mathcal{D}$ (solid-curves), and QO $\Omega$ (dashed-curve) for $\eta_{A B_{II}}$ with the same parameters as FIG. \ref{Fig1}.}
	\label{fig3}
\end{figure}
\end{minipage}
\end{widetext}

Now, we turn our attention to the influence of Bob's qubit within the GHS dilation spacetime in region-II. FIG. \ref{fig3} clarifies the behavior of QO and QD functions for varying values of $w$, $\alpha$, and $g$. A notable trend emerges as $D$ increases: the maximum values of both QO and QD rise until they converge towards a residual level. Generally, both functions originate from zero and gradually ascend with increasing $D$. FIG. \ref{fig3}. (a) shows that an augmentation in $g$ corresponds to an increase in the maximum values of both QO and QD, accompanied by a convergence in their behavior. FIG. \ref{fig3}. (b) reveals a complete congruence between certain QO states at elevated $\alpha$ values, with both QD and QO exhibiting similar patterns. This observation implies that QO effectively encapsulates quantum correlations. In contrast, FIG. \ref{fig3}. (c) indicates that higher QO and QD values are associated with lower frequencies of $w$, whereas elevated frequencies result in diminished values. These results underscore a reversal in the behavior of QO and QD within region-II, where the GHS spacetime parameters exert contrasting influences. While $w$ suppresses quantum behavior, $D$ promotes it. Conversely, the initial state parameters consistently enhance both functions as $g$ and $\alpha$ increase.

\begin{widetext}
\begin{minipage}{\linewidth}
\begin{figure}[H]
	\centering
	\includegraphics[scale=0.42,trim=00 00 00 00, clip]{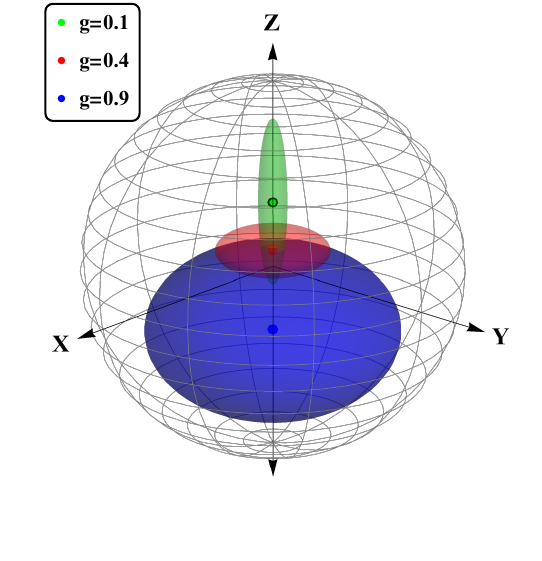} \quad
	\includegraphics[scale=0.42,trim=00 00 00 00, clip]{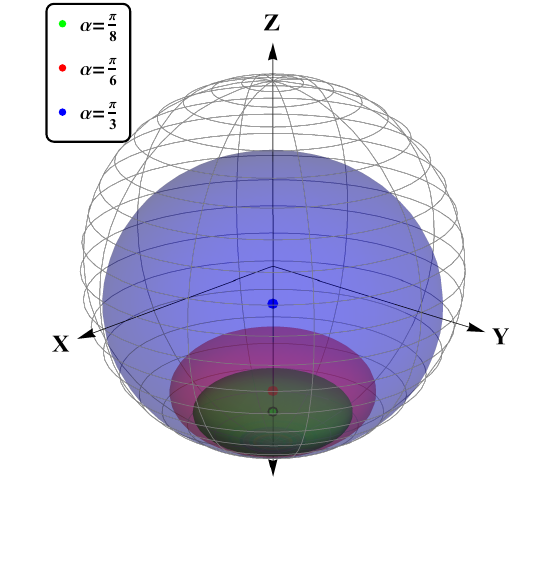} \quad
	\includegraphics[scale=0.42,trim=00 00 00 00, clip]{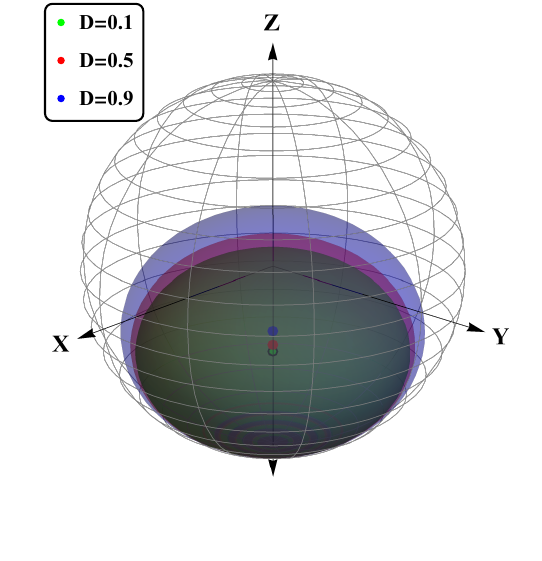} \quad
	\includegraphics[scale=0.42,trim=00 00 00 00, clip]{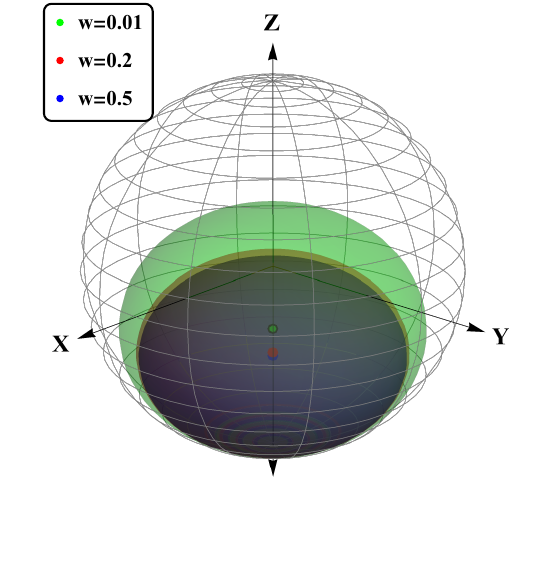} \put(-445,105){$
		\textbf{{\scriptsize (a)}}$} \put(-325,105){$\textbf{{\scriptsize (b)}}$} \put(-200,105){ $\textbf{{\scriptsize (c)}}$
	}\put(-80,105){ $\textbf{{\scriptsize (d)}}$}
	\caption{  Alice’s QSE $\varepsilon_{A}$ corresponding to $\eta_{AB_{II}}$ with the same parameters as FIG. \ref{Fig2}.}
	\label{Fig4}
\end{figure}
\end{minipage}
\end{widetext}
Regarding QSE in the region $II$, FIG. \ref{Fig4} offers a comprehensive analysis with the same parameters as FIG. \ref{Fig2}. The initial state parameters, $g$ and $\alpha$, as depicted in FIG. \ref{Fig4} (a) and (b), exhibit a comparable influence to that observed in FIG. \ref{Fig2}, albeit with a reduction in ellipsoid dimensions within the region $II$. Notably, FIG. \ref{Fig4} (c) and (d) showcase a contrasting trend in QSE behavior. The ellipsoid volume expands with increasing dilation parameter $D$ but remains constrained by the dimensions of the Bloch sphere. Furthermore, a decrease in Dirac field frequency correlates with a reduction in QSE volume.

\subsection{Bob's qubit in Region-I and II}

Considering the spacetime region, the reduced density operator in this region, $\eta_{B_{I}B_{II}}$ is

\begin{eqnarray}
	\eta_{B_{I}B_{II}} &=&(\frac{1-g}{2}+g\cos^{2}\alpha)\varepsilon_1^{2}|00\rangle\langle00|\nonumber\\&+&(\frac{1-g}{2}+g\cos^{2}\alpha)\varepsilon_1\varepsilon_2 (|00\rangle\langle11|+|11\rangle\langle00|)\nonumber\\&+& (\frac{1-g}{2}+g\sin^{2}\alpha)|10\rangle\langle10|\nonumber\\&+&(\frac{1-g}{2}+g\cos^{2}\alpha)\varepsilon_2^{2}|11\rangle\langle11|.\label{eq23}
\end{eqnarray}

\begin{widetext}
\begin{minipage}{\linewidth}
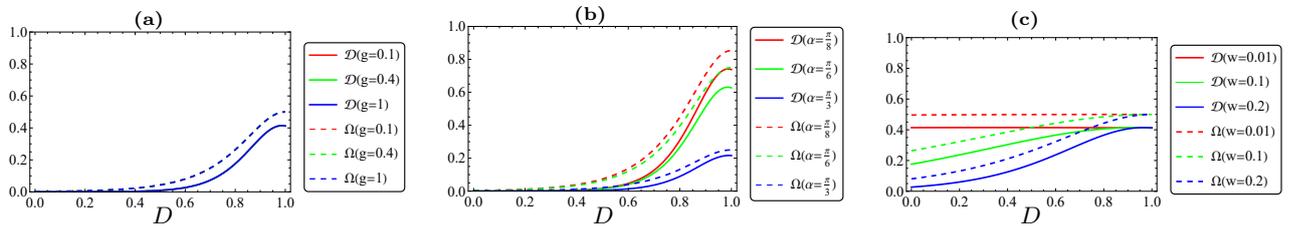
\begin{figure}[H]
	\centering
	\includegraphics[scale=0.42,trim=00 00 00 00, clip]{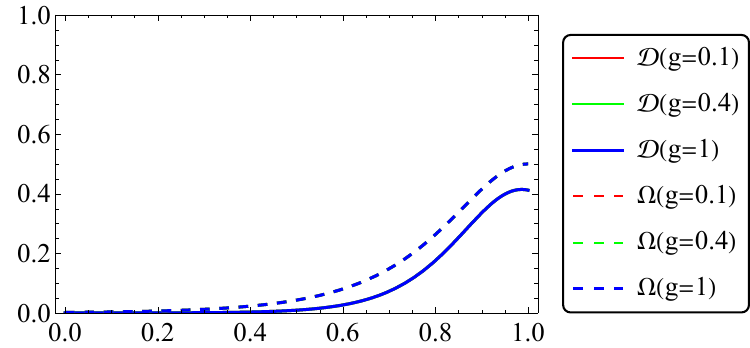} \quad
	\includegraphics[scale=0.42,trim=00 00 00 00, clip]{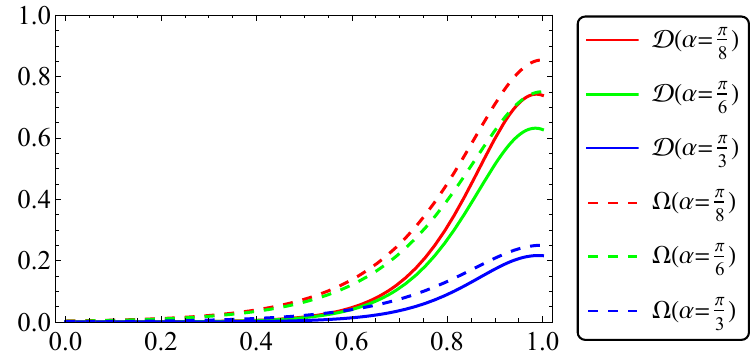} \quad
	\includegraphics[scale=0.42,trim=00 00 00 00, clip]{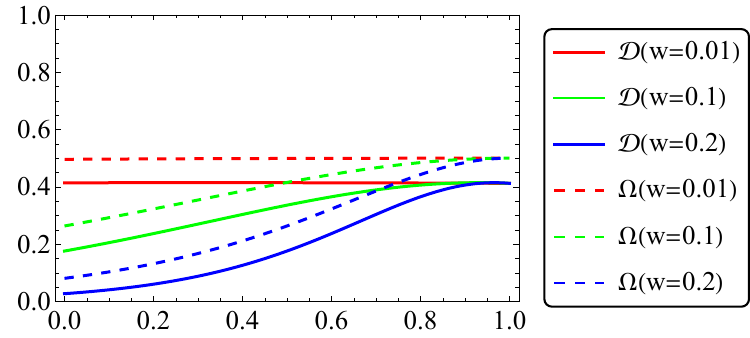}
	\put(-430,70){$\textbf{{\scriptsize (a)}}$} \put(-265,72){$\textbf{{\scriptsize (b)}}$} \put(-105,70){ $\textbf{{\scriptsize (c)}}$}  \put(-423,-5){${\scriptsize D}$} \put(-258,-5){${\scriptsize D}$} \put(-98,-5){${\scriptsize D}$} 
	\caption{ QD $\mathcal{D}$ (solid-curves), and QO $\Omega$ (dashed-curve) for $\eta_{B_{I}B_{II}}$ with the same parameters as FIG. \ref{Fig1}.}
	\label{Fig5}
\end{figure}
\end{minipage}
\end{widetext}

The behavior of the QO and QD functions for the state $\eta_{B_{I}B_{II}}$ was investigated under the same conditions as provided in FIG. \ref{Fig1}, with the results presented in FIG. \ref{Fig5}. Notably, these functions exhibit minimal dependence on the entanglement parameter, $g$. As illustrated in FIG. \ref{Fig5} (a), all curves coincide irrespective of $g$ values, with both functions increasing monotonically with $D$, while maintaining a consistent relationship. Conversely, the angle $\alpha$ significantly influences the functions' behavior. A decrease in entanglement, approaching a separable state, is correlated with an increase in both QO and QD values as $D$ grows, as depicted in FIG. \ref{Fig5} (b). A comparable trend is observed between FIG. \ref{fig3} (c) and \ref{Fig5} (c) when varying the parameter $w$, although both QO and QD exhibit lower magnitudes. Nevertheless, quantum correlations persist and intensify with increasing $D$.

\begin{widetext}
\begin{minipage}{\linewidth}
\begin{figure}[H]
	\centering
	\includegraphics[scale=0.42,trim=00 00 00 00, clip]{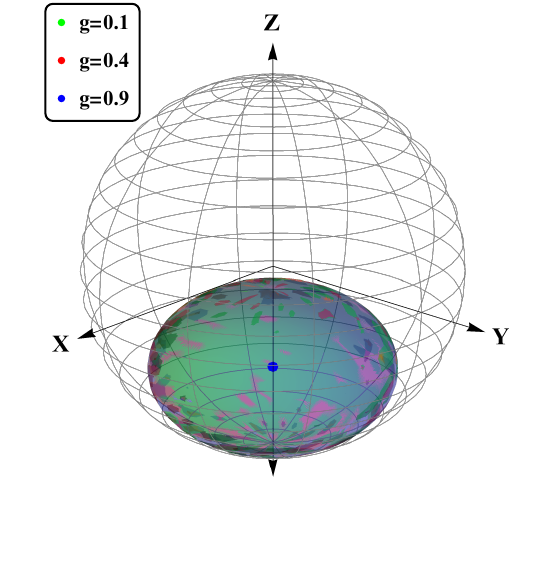} \quad
	\includegraphics[scale=0.42,trim=00 00 00 00, clip]{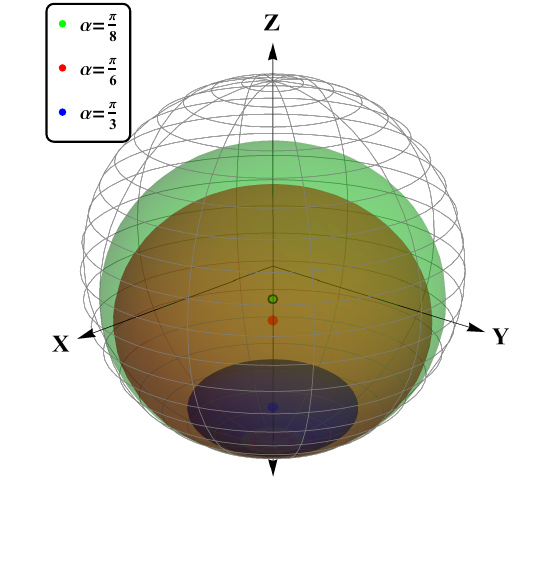} \quad
	\includegraphics[scale=0.42,trim=00 00 00 00, clip]{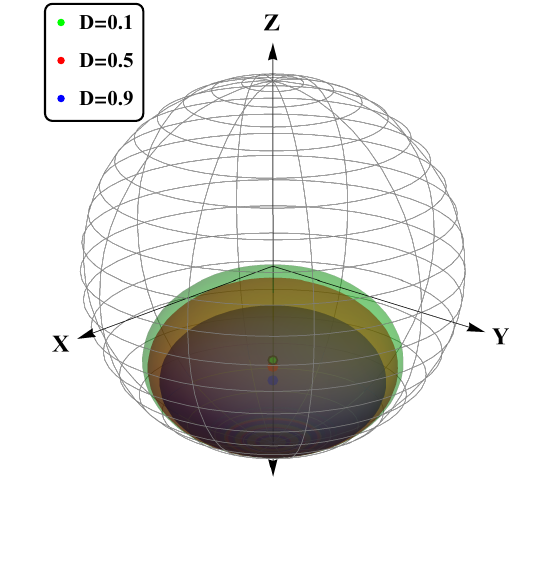} \quad
	\includegraphics[scale=0.42,trim=00 00 00 00, clip]{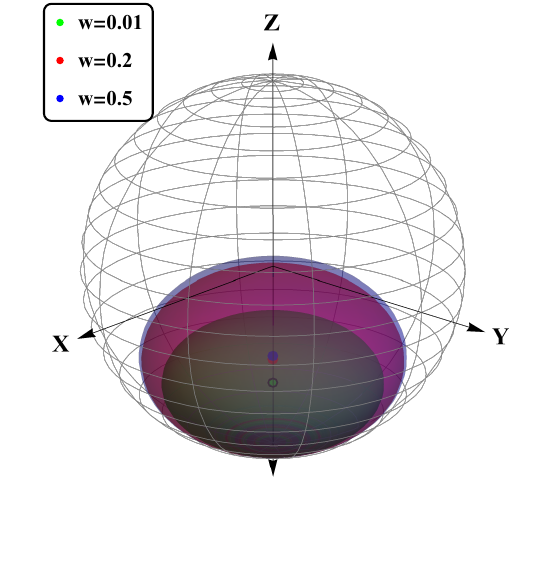} \put(-445,105){$
		\textbf{{\scriptsize (a)}}$} \put(-325,105){$\textbf{{\scriptsize (b)}}$} \put(-200,105){ $\textbf{{\scriptsize (c)}}$
	}\put(-80,105){ $\textbf{{\scriptsize (d)}}$}
	\caption{  Alice’s QSE $\varepsilon_{A}$ corresponding to $\eta_{B_{I}B_{II}}$ with the same parameters as FIG \ref{Fig2}.}
	\label{Fig6}
\end{figure}
\end{minipage}
\end{widetext}
FIG. \ref{Fig6} depicts the behavior of QSE between Bob's qubit in regions $I$ and $II$. A notable consistency in behavior is observed across varying values of the parameter $g$, with the ellipsoid's center consistently positioned on the negative Z-axis, as verified by FIG. \ref{Fig5}. (a). Moreover, an increase in $\alpha$ correlates with a diminution in QS ellipsoid size and a concomitant downward shift of its center along the Z-axis. While modifications to parameters $D$ or $w$ induce subtle alterations in ellipsoid dimensions, a slight expansion of the QSE is observed with decreasing dilation parameter or increasing Dirac field frequency.

\section{Concluding remarks \label{sec5}}
In this paper, we have investigated QO, QD, and the QSE of a Gisin state under GHS dilation spacetime. The Gisin state, comprising an entangled state and a mixed state, had been selected as the initial state. It is assumed that Alice remained stationary in an asymptotically flat region, and Bob, located near the event horizon in GHS dilation spacetime. The three quantum quantifiers have been studied across three scenarios: Alice’s qubit stationary with Bob’s qubit in GHS dilation spacetime, Alice’s qubit stationary with Bob’s qubit in anti-GHS dilation spacetime, and the correlation between Bob’s qubit in GHS and anti-GHS dilation spacetime. 

Our results have indicated that the physical accessibility of correlation depended upon the region of detection. For the first scenario, QO and QD have decreased as the dilation parameter increased, while increasing with the Dirac frequency. The mixed state component appeared to have signified a decoherence state. An elevated mixing parameter had inversely correlated with quantum correlation. The interplay between the initial state angle and the mixing parameter might have influenced the augmentation or reduction of quantum correlation. In the second and third scenarios, the effect had been similar due to the changing dilation parameter, where QD and QO increased with dilation. Increasing the Dirac frequency led to decreasing QD and QO, although the maximum values of the two functions had been larger in the second scenario compared to the third. For QSE, initial state parameters had been similarly affected in the first and second scenarios, with a decreased mixing parameter reducing the QSE, while angle settings worked to achieve a perpendicular state approaching the size of the Bloch sphere. Conversely, in the third scenario, the QSE did not depend on the mixing parameter, and decreasing angle settings (reaching the separable state) had sufficed to attain the size of the Bloch sphere. These results have been consistent with previous studies of entanglement in spacetime \cite{metwally2013usefulness,he2016measurement,ming2024nonlocal}. Another important observation for QSE, when the Dirac frequency was increased to a significant value in the first scenario, the QSE matched the size of the Bloch sphere, enabling the possibility of steerability despite a large dilation parameter.

\bibliography{references}
\bibliographystyle{unsrt}

\end{document}